# SAR ALTIMETRY APPLICATIONS OVER WATER

**Cristina Martin-Puig[(1)], Jose Marquez[(1)], Giulio Ruffini[(1)], R. Keith Raney[(2)], Jérôme Benveniste[(3)]**

*(1) Starlab Barcelona S.L., Camí de l'Observatori s/n 08830 Barcelona Spain; cristina.martin@starlab.es
(2)Johns Hopkins University Applied Physics Laboratory, Laurel MD 20723-6099 USA; keith.raney@jhuapl.edu
(3) European Space Agency ESRIN, Via Galileo Galilei I-00044 Frascati, Italy; Jerome.Benveniste@esa.int
Jan 2008*

**ABSTRACT/RESUME**

The application of Synthetic Aperture Radar (SAR) techniques to classical radar altimetry offers the potential for greatly improved Earth surface mapping. This paper provides an overview of the progress of SAMOSA, *Development of SAR Altimetry Studies and Applications over Ocean, Coastal zones and Inland waters*, an on-going ESA-funded project. The main objective of SAMOSA is to better quantify the improvement of SAR altimetry over conventional altimetry on water surfaces. More specifically, one of the tasks focuses on the reduction of SAR mode data to pulse-limited altimeter data, and a theoretical modelling to characterize the expected gain between high Pulse Repetition Frequency (PRF) reduced SAR mode data and low PRF classical Low-Resolution Mode (LRM) data. To this end, theoretical modelling using the Crámer-Rao bound (CRB) will be used and the results will be compared to previous theoretical estimates [7], using an analysis akin to that in [8].

## 1 INTRODUCTION

With a launch scheduled in 2009, CryoSat-2's main instrument, SIRAL, will be the first altimeter to implement SAR techniques to classical radar altimetry. In addition to cryospheric applications, the data produced by such an altimeter would be of great interest to the hydrosphere and oceanographic communities since it will allow quantitative assessment of expected enhanced altimetric capabilities in coastal monitoring, ocean floor topography, gravity field and inland water monitoring.

SAR altimetry was first described as Delay/Doppler Radar Altimeter in 1998 [1]. Its key innovation is the addition of along track processing for increased resolution and multi-look processing. This technique requires echo delay compensation, analogous to range cell migration correction in conventional SAR processing [2]. Due to this innovation, spatial resolution is increased in the along-track dimension. In turn, this allows for accumulation of more statistically independent looks for each scattering area, leading to better speckle reduction, hence finer precision of altimetric measurements.

Under the framework of the ESA's SAMOSA project, the improvement of SAR altimetry when compared to conventional altimetry over water surfaces is to be quantified. SAMOSA kicked off in September 2007, and its expected duration is fifteen months. The project, led by Satellite Observing Systems (SOS, UK), is composed of four additional scientific partners with experience in space oceanography: The Danish National Space Centre (NDSC, Denmark), De Montfort University (DMU, UK), The National Oceanography Centre (NOCS, UK) and Starlab Barcelona S.L: (STARLAB, Spain). This consortium, with the external participation of Dr. R.K. Raney (Johns Hopkins University, USA), is analysing the potentials of advanced SAR Altimetry over water surfaces.

SAMOSA is subdivided in eight different tasks:

- Task 1: *State of the art assessment*
- Task 2: *Range error as a function of ocean surface*
- Task 3: *Recovery of short wavelength geophysical signals and short spatial scale sea surface slope signals*
- Task 4: *SAR Altimetry echo over water*
- Task 5: *New re-tracking method over water*
- Task 6: *Improvement of capabilities for coastal zone, estuaries, rivers and lakes*
- Task 7: *Assessment of RA-2 individual echoes over water*
- Task 8: *Validation using airborne ASIRAS data*

The tasks are not sequentially sorted. Some tasks may take place in parallel. Task 1 is already completed and the project current activities are Task 2 and Task 3.

This paper presents the theoretical progress of SAMOSA Task 2. Task 2 aims to do a scientific study of the potential improved capabilities of the CryoSat SAR data over ocean when compared to conventional altimeters. Assuming the lack of a re-tracker for the SAR altimeter observing water surfaces (work to be done in task 5 of this project), the approach for the performance analysis is based on the use of SAR mode data to emulate classical altimeter data (aka. Low Resolution Mode, LRM, for SIRAL), and compare the theoretical performance of the emulated and real LRM data with a conventional altimeter re-tracker. In effect, SAR mode data will be transformed such that it emulates LRM data, and the result of the transformation will be hereafter referred as *Reduced SAR data*. The theoretical analysis involved in the performance estimation is detailed in this document. Numerical results are not presented in this paper, and may be found in future deliverables of the SAMOSA project.

## 2 CRYOSAT ACQUISITION MODES

CryoSat's SIRAL altimeter has three operating modes: the LRM, the SAR mode and the inSAR mode [3]. The first two are of main interest to understand the work presented in the forthcoming sections.

In LRM the altimeter performs as a conventional pulse limited altimeter. This mode operates at a pulse repetition frequency (PRF) of 1970 Hz. This PRF is low enough to ensure that the echoes are decorrelated. Therefore, the echoes received may be incoherently added to reduce speckle noise by a $1/\sqrt{M}$ factor, where M is the number of averaged pulses in the selected time interval.

In SAR mode the pulses are transmitted in bursts. Correlation between echoes is desired [1], thus in this mode the PRF within a burst is higher than the LRM one and equal to 17.8 KHz [3]. Each burst transmits 64 pulses at this PRF. After transmission the altimeter waits for the returns and transmits again the next burst. Therefore, there is not only an 'intra-burst' PRF, but also a burst repetition frequency (BRF), which according to SIRAL specifications is of 85.7 Hz [3].

## 2 CRYOSAT DATA PRODUCTS

The previous acquisition modes will provide different data products: level 1 or full bit rate data (FBR), level 1b or multi-looked waveform data, and level 2 for evaluation or geophysical products. This paper is only addressing FBR data for LRM and SAR mode.

For LRM FBR the echoes are incoherently added on board prior to altimetry. In this mode, FBR data consists of multi-looked echoes at a rate of approximately 20Hz.

Unlike LRM, in SAR mode the echoes must be SAR processed before incoherently multi-looking. The SAR processing is mostly done on ground. SAR FBR data consists of complex waveforms (I and Q components) telemetered before the IFFT block (see Figure 1).

## 2 REDUCED SAR MODE DATA

### 2.1. Description

Reduced SAR mode data is the SAR FBR data transformed such that it emulates LRM data.

### 2.2 Motivation for Reduced SAR

CryoSat operational modes are exclusive. Reduced SAR mode offers two types of results from the same FBR acquisition: emulation of a conventional altimeter and a SAR altimeter performance. This allows for a quantitative comparison of the measurement precision over identical sea state.

### 2.3 Methodology to Achieve Reduced SAR mode

SIRAL LRM and SAR modes both transmit pulses of identical pulse length. The main difference between these modes is the PRF, and its associated effects. In SAR mode, pulse to pulse correlation is a consequence of its high PRF, while pulse to pulse correlation is not present in LRM, nor desired. The approach presented in this section to convert SAR mode data into LRM is mainly focused on the PRF difference of both modes.

To reduce SAR mode data to emulate LRM, a set of *n* SAR complex echoes needs first to be coherently added (a.k.a. pre-summing, see Figure 3), since they are correlated and incoherent summation is not possible. Then, the resulting waveforms (complex) are processed as if they were the input to a conventional altimeter (input to the IFFT block of Figure 1).

Figure 1 shows a high-level block diagram of a conventional altimeter.

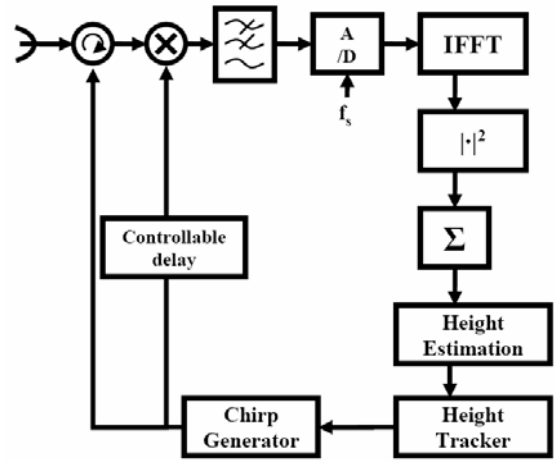

*Figure 1: Conventional altimeter high level block diagram*

Figure 2 provides a high-level block diagram of a SAR altimeter. Highlighted in green the new processing blocks added.

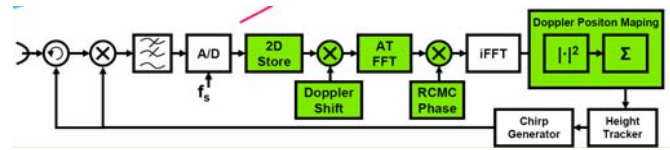

*Figure 2: SAR altimeter block diagram*

SAR FBR data is provided before SAR processing. It is provided before the 2D store block. Comparing Figure 1 and Figure 2, if we do the coherent pre-sum of *n* pulses and process the resulting complex waveforms as if they were input to a conventional altimeter, the processes missing to emulate a LRM would be: the IFFT and the detection (power transformation). The final waveforms achieved after doing the previous procedure will be power waveforms, hereafter referred as pseudo-LRM.

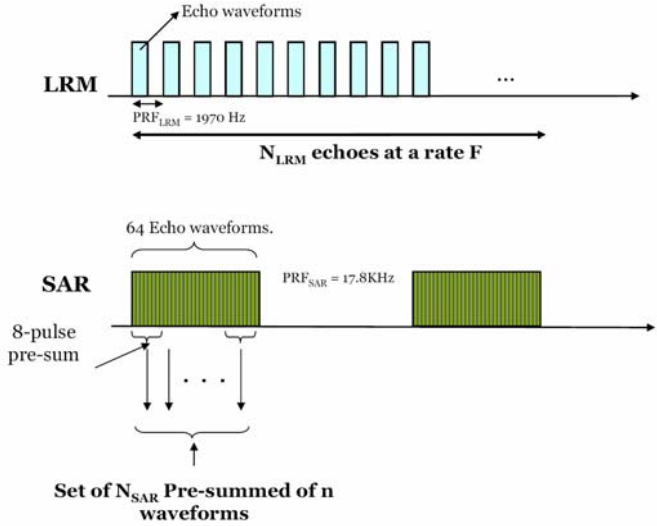

*Figure 3: Reduced SAR graphical representation of pulse coherent pre-sum concept*

After IFFT and detection it would be convenient to have decorrelated waveforms to apply incoherent summation and reduce speckle, as it is done with LRM FBR data. The main issue at this stage is to identify the number of pulses to be coherently added *n*. This number should be as small as possible but decorrelation of the resulting pre-summed pulses needs to be ensured.

The number *n* can be calculated from the PRF ratio of both modes LRM and SAR mode. Doing such procedure we intend to emulate the LRM PRF with the SAR FBR data. The closes integer (floor) of

$$n = \frac{PRF_{SAR}}{PRF_{LRM}}$$  **Eq. 1**

is 9. However, since the burst comprises 64 pulses a better choice of *n* would be 8, resulting into a sequence of complex waveforms at PRF slightly higher that the LRM. Accounting for the 64 pulses per burst, if groups of 8 are done, this leads to 8 pseudo-LRM waveforms per burst after IFFT and detection, with PRF of 2.22KHz.

SIRAL is the first altimeter with "noncoherent altimeter" mode (LRM) and "coherent altimeter" mode (SAR). Pulse-to-pulse correlation and decorrelation is measured by the analysis of the PRF used in each mode. To ensure coherence of pulses in SAR mode, the PRF must be higher than the rate that would correspond to two samples per along-track movement equal to the diameter of the radar antenna. Given CryoSat's specifications the lower bound PRF to ensure pulse to pulse correlation is approximately 15KHz. Therefore, CryoSat choice of 17.8KHz will satisfy pulse-to-pulse correlation and enable coherent processing.

On the other hand, pulse-to-pulse decorrelation is not desired in LRM. In 1982 [4] a method to measure the upper bound of the decorrelation PRF was presented. This method, based on the principle of Van Cittert-Zernike, states that the decorrelation PRF is obtained by dividing the spacecraft velocity by the decorrelation distance and correcting for the curvature of the earth:

$$PRF = \frac{v}{d\left(\frac{R_e + h}{R_e}\right)}$$  **Eq. 2**

$$d = \frac{0.305 \lambda h}{r}$$  **Eq. 3**

Here $\lambda$ is the radar wavelength, $h$ the altimeter altitude, and $r$ the radius of the circular uniformly illuminated area. Note that for non flat surfaces (e.g. ocean surface) r will increase with the SWH [9] by:

$$r = \left[\frac{(c\tau + 2H_{1/3})h}{1 + h/R_e}\right]^{1/2},$$  **Eq. 4**

with c the speed of light, $\tau$ the lag between the leading and trailing edges of the pulse and $H_{1/3}$ the SWH. **Eq. 4** combined with **Eq. 2**, **Eq. 3** result into an increase of the useful PRF with increasing SWH.

Figure 4 shows the upper bound on PRF as a function of the SWH.

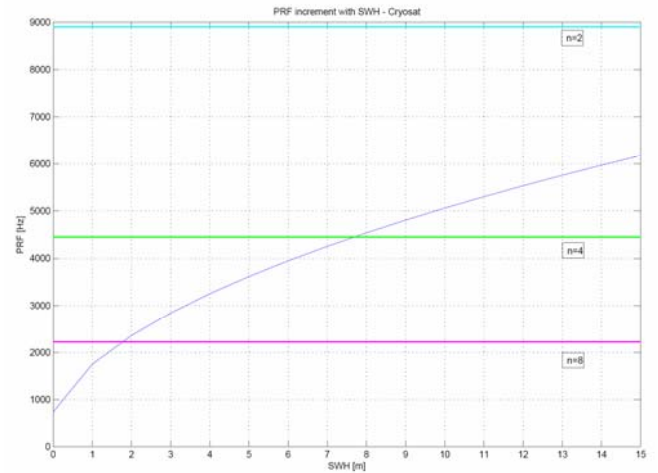

***Figure 4** PRF as a function of SWH; added lines of PRF values achieved by adding coherently n=2, n=4 and n=8 SAR FBR. Figure attached at the end with better resolution.*

For n equal to 8, the decorrelated PRF is below the upper bound for a SWH of 2 m. For n integer of 64 and lower than 8, like 4 or 2, our upper bound will correspond to higher SWHs, specially for a choice of *n*=2. The choice of n=4 would be still acceptable for very rough seas state. However, for this study *n=8* for a SWH of 2m is very reasonable.

To determine the precision of reduced SAR compared to a conventional altimeter we will make use of estimation theory, i.e., we will analyse the Cramér-Rao bound.

## 3 CRAMÉR-RAO BOUND (CRB)

The CRB provides the best achievable performance for an estimation process in which the stochastic nature of the observation can be described by a probability distribution function (pdf) [5].

Provided with a stochastic observation *X* (e.g., the complex waveform, a vector) and a pdf(*X*) which is a function of the parameter to estimate, *θ* (e.g., the two way travel delay), our interest is to investigate our precision (variance) in estimating *θ* from the measurement *X*, given the functional relation pdf(*X*) = *p*(*X*;*θ*). The CRB is formally defined by

$$CRB = \min\left[\sqrt{E[(\hat{\theta}-\theta)^2]}\right] \quad \textbf{Eq. 5}$$

for unbiased estimators of *θ*. If $p(X;\theta)$ satisfies

$$E\left[\frac{d\ln(p(X;\theta))}{d\theta}\right]=0 \quad \forall \theta \quad \textbf{Eq. 6}$$

then,

$$\text{var}(\hat{\theta}) \geq CRB^2 = \frac{1}{-E\left[\frac{\partial^2 \ln p(X;\theta)}{\partial \theta^2}\right]} \quad \textbf{Eq. 7}$$

Focusing on complex, vectorial Gaussian-distributed signals, the PDF is given by:

$$p(X,\theta) = \frac{1}{\pi^{card(X)}|\Gamma|}\exp\left[-(X-E[X])^+ \cdot \Gamma^{-1} \cdot (X-E[X])\right] \quad \textbf{Eq. 8}$$

with $\Gamma_{ij} = E\left[(X_i - E[X_i])(X_j - E[X_j])^*\right]$ the covariance matrix of the complex signal vector *X* respectively. Provided with the fact that in our the waveform samples are decorrelated, the CRB expression simplifies to

$$CRB^{-2} = \sum_i \frac{1}{\Gamma_i}\frac{\partial \Gamma_i}{\partial \theta} + \frac{2}{\Gamma_i}\left(\frac{\partial E[X_i]}{\partial \theta}\right)^2. \quad \textbf{Eq. 9}$$

**Eq. 9** is the starting point of the theoretical precision analysis presented in the forthcoming sections.

## 4 CONVENTIONAL ALTIMETER COMPLEX WAVEFORM

The CRB can be applied to the waveform after incoherent averaging or to the complex waveform after IFFT previous to detection. The analysis described in this section focuses on the application of CRB for range precision estimation applied to the complex waveform after the IFFT block (see Figure 1).

The complex detected waveform noise is associated to two stochastic phenomena: the radar electric field scattered over the ocean's random surface, and thermal noise originating from the scene and instrument. The waveform can be modelled by

$$X = \alpha S_R \otimes S_d + \sigma N \otimes S_d = \alpha u + \sigma N \otimes S_d, \quad \textbf{Eq. 10}$$

where $S_R$ is the scattered electric field and *u* the electric field convolved with the de-ramping signal $S_d$. *N* is the thermal noise, which will be assumed to be complex, vectorial, with real and imaginary components both Gaussian, zero mean and, without loss of generality, with standard deviation 1. Therefore, the complex detected waveform will also be Gaussian, with

$$E[X_i] = 0, \text{ and} \quad \textbf{Eq. 11}$$

$$\Gamma_{ij} = \alpha^2 E[u_i u_j^*] + \frac{\pi}{4-\pi}\frac{1}{SNR_V^2}\chi(\theta;H_{1/3}) \cdot \quad \textbf{Eq. 12}$$

Here is $\chi(\theta;H_{1/3})$ the radar ambiguity function dependent on the delay time and significant wave height (SWH), and equal to $\chi(\theta;H_{1/3}) = S \otimes S_d$, where *S* is the transmitted chirp signal. $SNR_V$ is the thermal signal to noise ratio (voltage) of the waveform.

From the previous results, **Eq. 9** simplifies to:

$$CRB^{-2} = \sum_i \frac{1}{\Gamma_i}\frac{\partial \Gamma_i}{\partial \theta}, \quad \textbf{Eq. 13}$$

The next step is the evaluation of the scattered filtered field $E[u_i u_j^*]$. The critical feature of our analysis is the waveform leading edge, since it relates to the delay time to mean sea level (MSL; $\theta$) and the SWH [6].

The leading edge is obtained by integration of sea-surface scatterers in the vicinity of the specular point. In this regime and from space, the signal covariance is largely dominated by the radar ambiguity function, thus the antenna pattern and the glistening zones can be considered to be larger than the first pulse-limited footprint area (area defined by $t = t_{crest} + \tau + \frac{2H_{1/3}}{c}$; $\tau$ pulse length). In addition, SIRAL observation will be nadir looking. Accounting for the previous information, $u$ can be expressed using a simple electromagnetic model (physical optics):

$$u = \int_{surface} G(\vec{\rho}) \chi[\theta; H_{1/3}] \cdot \frac{e^{jkr}}{r} d\vec{\rho}. \qquad \textbf{Eq. 14}$$

Here $G(\vec{\rho})$ is the antenna footprint over the surface, $\vec{\rho}$ the horizontal position vector, $r$ the distance between the scatterers and the receptor and $k$ the wave number.

The previous theory applies equally to LRM and reduced SAR mode in terms of CRB. After the IFFT the complex waveform is transformed into power by calculating its absolute value squared, and when such transformation is completed the waveforms are incoherently added.

After incoherent integration of $M$ different waveforms the SNR voltage can be expressed as

$$SNR = \sqrt{M} \cdot SNR_V. \qquad \textbf{Eq. 15}$$

Thus, if $M$ differs from the different modes so will the precision of the system.

Considering a time window $T$ or a set of multi-looked echoes at a rate of $F=1/T$, which preferably should be a multiple of the BRF, $T = \beta \cdot 11.7 ms$ being $\beta$ an integer number greater than zero, the number of LRM echoes received in this time window is equal to $M_{LRM} = T \cdot PRF_{LRM}$. The number of reduced SAR mode achievable in the same time window is equal to $M_{\text{Re}duced\_SAR} = \beta \cdot 8$, if we define $n=8$ as explained in section 2.3. Considering the above, if a comparison between $SNR_{LRM}$ versus $SNR_{\text{Re}duced\_SAR}$ is carried out, the result is a -2.29dB degradation of Reduced SAR mode compared to LRM.

## 5 CONCLUSIONS AND FUTURE WORK

The Reduced SAR mode will allow quantitative comparison of emulated LRM and SAR mode over identical sea states in terms of altimetric precision.

A theoretical degradation in SNR (voltage) of -2.29dB has been estimated. This degradation may still be considered acceptable for specific altimetric observations. Means are being explored that would eliminate the -2.29dB degradation through additional signal processing methods.

Software to reduce SAR mode in Reduced SAR has been developed to numerically simulate range precision with a conventional altimeter re-tracker. Input data to the software will be generated by CRYMPS (SIRAL data simulator). The work to be carried out with numerical simulations will further analyse the results provided here and in [7].

## 6 ACKNOWLEDGEMENTS

The authors would like to acknowledge the SAMOSA team (ESRIN Contract No. 20698/07/I-LG) the CRYMPS team (MSSL), and Maria Milagro (Serco support to ESA-ESRIN)

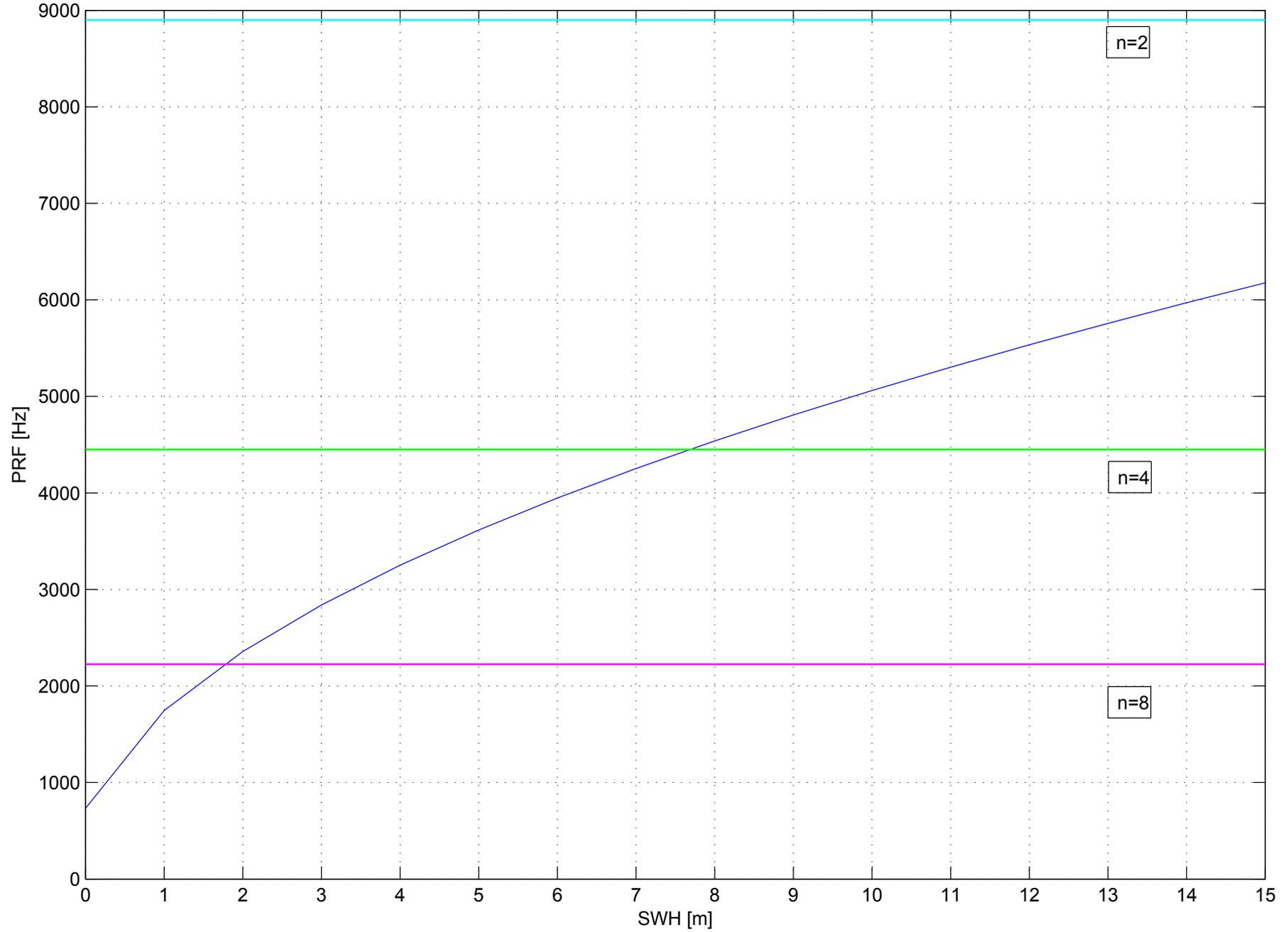